# Performance Analysis of FDDI Token Ring Networks: Effect of Parameters and Guidelines for Setting TTRT [1]

Raj Jain
Digital Equipment Corp.
550 King St. (LKG 1-2/A19)
Littleton, MA 01460
Internet: Jain@Erlang.enet.DEC.Com

**Abstract**

The performance of Fiber-Distributed Data Interface (FDDI) depends upon several workload parameters; for example; the arrival pattern, frame size, and configuration parameters, such as the number of stations on the ring, extent of the ring, and number of stations that are waiting to transmit. In addition, the performance is affected by a parameter called the Target Token Rotation Time (**TTRT**), which can be controlled by the network manager. We considered the effect of TTRT on various performance metrics for different ring configurations, and concluded that a TTRT value of 8 ms provides a good performance over a wide range of configurations and workloads.

# 1 Introduction

Fiber-Distributed Data Interface (FDDI) is a 100-Mbps local-area network standard being developed by the American National Standards Institute also known as ANSI [1]. The standard allows up to 500 stations to communicate via fiber optic cables using a timed-token access protocol. Normal data traffic as well as time constrained traffic such as voice, video, and real-time applications are supported. All major computer vendors, communications vendors, and integrated circuit manufacturers are offering products supporting this standard.

Unlike the token access protocol of IEEE 802.5, FDDI uses a timed-token access protocol that allows both synchronous and asynchronous traffic simultaneously. The maximum access delay, the time between successive transmission opportunities, is bounded for both synchronous and asynchronous traffic. Although the maximum access delay for the synchronous traffic is short, that for asynchronous traffic can be long depending upon the network configuration and load. As is shown later, unless care is taken, the access delay can be as long as 165 seconds. This means that a station wanting to transmit asynchronous traffic may not get a usable token for 165 seconds. Such long access delays are clearly not desirable and can be avoided by proper setting of the network parameters and configurations. TTRT is one such parameter. The effect of this parameter on various performance metrics was investigated and guidelines for setting its value were developed.

[1] This is a modified version of the paper [5] presented at ACM SIGCOMM, Philadelphia, PA, September 1990.

## 2 Timed-Token Access Method

A token access method, for example, the one used on IEEE 802.5, works as follows. A token is circulated around the ring. Whenever a station wants to transmit, it waits for the token arrival. Upon receiving a token, it can transmit for a fixed interval called the Token Holding Time (**THT**). After the transmission, the station either releases the token immediately or after the arrival of all the frames it transmitted. Using this scheme, a station on an $n$ station ring may have to wait as long as an $n\times$THT interval to receive a token. This may be unacceptable for some applications if $n$ or THT is large. For example, for voice traffic and real-time applications, this interval may be limited to the 10-20 ms range. Using the token access method severely limits the number of stations on the rings.

The timed-token access method, invented by Grow [2], solves this problem by ensuring that all stations on the ring agree to a 'target' token rotation time and limit their transmissions to meet this target as much as possible. There are two modes of transmission: synchronous and asynchronous. Time-constrained applications such as voice and real-time traffic use the synchronous mode. Traffic that does not have time constraints uses the asynchronous mode. The synchronous traffic can be transmitted by a station whenever it receives a token. The total time of transmission per opportunity is, however, short, and it is allocated at the ring initialization. The asynchronous traffic can be transmitted only if the token rotation time is less than the target.

The basic algorithm for the asynchronous traffic is as follows. Each station on the ring measures the time since it last received the token. The time interval between two successive receptions of the token by a station is called the Token Rotation Time (TRT). On a token arrival, if a station wants to transmit, it computes a Token Holding Time (THT):

$$THT = TTRT - TRT$$

Here, TTRT is the *target* token rotation time as agreed by all stations on the ring. If THT is positive, the station can transmit for this interval. At the end of transmission, it releases the token. If a station does not use the entire THT allowed, other stations on the ring can use the remaining time by using the same algorithm

Notice that even though the stations attempt to keep TRT below the target, they do not always achieve their goal. It is possible for TRT to exceed the target by as much as the sum of all synchronous transmission-time allocations. Actually, the synchronous time allocations are limited so that their sum is less than TTRT. This ensures that the TRT is always less than two times TTRT.

# 3 Performance Parameters

The performance of any system depends upon the workload as well as the system parameters. There are two kinds of parameters: fixed and user settable. Fixed parameters are those that the network manager has no control over. These parameters vary from one ring to the next. Examples of fixed parameters are cable length and number of stations. It is important to study performance with respect to these parameters since, if it is found that performance is sensitive to these, a different guideline may be used for each set of fixed parameters. The settable parameters, which can be set by the network manager or the individual station manager, include various timer values. Most of these timers affect the reliability of the ring and the time to detect malfunction. The key parameters that affect a performance are the TTRT and the synchronous time allocations.

The workload also has a significant impact on system performance. One set of parameters may be preferable for one workload but not for another. The key parameters for the workload are: the number of *active* stations and the load per station. By active we mean stations that are either transmitting or waiting to transmit on the ring. There may be a large number of stations on the ring, but only a few of these are generally active at any given time. The active stations include those that have frames to transmit and are waiting for the access right, that is, for a usable token to arrive along with the currently transmitting station, if any.

In this paper, the performance has been studied under asynchronous traffic only. The presence of synchronous traffic will further restrict the choice of TTRT.

# 4 Performance Metrics

The quality of service provided by a system is measured by its productivity and responsiveness [3]. For FDDI, productivity is measured by its throughput and responsiveness is measured by the response time and access delay. The response time is defined as the time between the arrival of a frame and the completion of its transmission ("first-bit in" to the "last-bit out"). Since this includes queueing delay, it is a meaningful metric only if the ring is not saturated. At loads near or above capacity, the response time reaches infinity and does not offer any information. With these loads, the access delay, which is defined as the time to get a usable token ("want-token" to "get-token" interval) is more meaningful.

The productivity metric that the network manager may be concerned with is the total throughput of the ring in Mbps. Over any reasonable interval, the throughput is equal to the load. That is, if the load on the ring is 40 Mbps, the throughput is also 40 Mbps. This, of course, does not hold if the load is high. For example, if there are three stations on the ring, each with a 100 Mbps load, the total arrival rate is 300 Mbps and the throughput is obviously much less. Thus, the key metric is not the throughput under low load but the maximum obtainable throughput under high load. This latter quantity is also called the *usable bandwidth* of the network. The ratio of the usable bandwidth to nominal bandwidth (100 Mbps for FDDI) is defined as the *efficiency*. Thus, if for a given set of network and workload parameters, the usable bandwidth on FDDI is never more than 90 Mbps, the efficiency is 90% for that set of parameters.

Another metric that is of interest for a shared resource, such as FDDI, is the **fairness** with which the resource is allocated. Fairness is particularly important under heavy load. However, the FDDI protocols have been shown to be fair provided the priority levels are not implemented. Given a heavy load, the asynchronous bandwidth is equally allocated to all active stations. In the case of multiple priority implementation, Dykeman and Bux [4] have shown that the protocol is *not* fair in the sense that it is possible for two stations with the same priority and same load to get different throughput depending upon their location. Low-priority stations closer to high-priority stations may get a better service than those further down stream. A single priority implementation is assumed here to keep the analysis simple. Such implementations have no fairness problems and, therefore, this metric will not be of concern anymore in this paper.

Two different methods have been used to analyze performance: simulation and analytical modeling. Analytical modeling is used to compute the efficiency and access delay under heavy load. A simulation model is used to analyze the response time at loads below the usable bandwidth. The response time does depend upon the arrival pattern and, therefore, a particular workload is used, which is described in the next section.

## 5 Simulation Workload

The workload used in the simulations was based on an actual measurement of traffic at a customer site. The chief application at this site was Warehouse

Inventory Control (**WIC**). Hence, the workload is called the "WIC Workload." Measurements on networks have shown that when a station wants to transmit, it generally transmits not one frame, but a burst of frames. This was found to be true in WIC workloads as well. Therefore, a "bursty Poisson" arrival pattern is used in the simulation model. The interburst time used was 1 milliseconds and each burst consisted of five frames. The frames had only two sizes: 65% of the frames were small (100 bytes) and 35% were large (512 bytes). A simple calculation shows that this workload constitutes a total load of 1.23 Mbps. Forty stations, each executing this load, would load an FDDI to 50% utilization. Higher load levels can be obtained either by reducing the interburst time or by increasing the number of stations.[2]

## 6 A Simple Analytical Model

The access delay and efficiency are meaningful only under heavy load and, therefore, it is assumed that there are $n$ active stations and that each one has enough frames to keep the FDDI fully loaded.

For an FDDI network with a ring latency of $D$ and a TTRT value of $T$, the efficiency and maximum access delay are [5]:

$$\text{Efficiency} = \frac{n(T - D)}{nT + D} \quad (1)$$

$$\text{Maximum access delay} = (n-1)T + 2D \quad (2)$$

[2] The measured inter-burst time was approximately 8 milliseconds. It was scaled down to represent more powerful processors and to get meaningful results while keeping the number of stations in the simulation small.

Equations 1 and 2 can be used to compute the maximum access delay and the efficiency for any given FDDI ring configuration. For example, consider a ring with 16 stations and a total fiber length of 20 km.[3] Light waves travel along the fiber at a speed of 5.085 $\mu$s/km. The station delay, the delay between receiving a bit and repeating it on the transmitter side, is of the order of 1 $\mu$s per station. The ring latency can, therefore, be computed as follows:

$$\begin{aligned} \text{Ring Latency } D &= (20 \text{ km}) \times (5.085\ \mu\text{s/km}) \\ &\quad + (16 \text{ stations}) \times (1\ \mu\text{s/station}) \\ &= 0.12 \text{ ms} \end{aligned}$$

Assuming a TTRT of 5 ms, and all 16 active stations, the efficiency and maximum access delay are:

$$\text{Efficiency} = \frac{16(5 - 0.12)}{16 \times 5 + 0.12} = 97.5\%$$

$$\begin{aligned} \text{Maximum access delay} &= (16 - 1) \times 5 + 2 \times 0.12 \\ &= 75.24 \text{ ms} \end{aligned}$$

Thus, on this ring the maximum possible throughput is 97.5 Mbps. If the load is more than this for any substantial length of time, the queues will build up, the response time will become very long, and the stations may start dropping the frames. The maximum access delay is 75.24 ms, that is, it is possible for asynchronous stations to take as long as 75.24 ms to get a usable token.

---

[3] Using a two-fiber cable, this would correspond to a cable length of 10 km.

The key advantage of this model is its simplicity, which allows us to immediately see the effect of various parameters on the performance. With only one active station, which is usually the case, the efficiency is:

$$\text{Efficiency with one active station} = \frac{T-D}{T+D}$$

As the number of active stations increases, the efficiency increases. With a very large number of stations ($n=\infty$), the efficiency is:

$$\text{Maximum efficiency} = 1 - \frac{D}{T}$$

This formula is easy to remember and can be used for "back-of-the-envelop" calculation of the FDDI performance. This special case has already been presented by Ulm[6].

Equation 2 also indicates that the maximum access delay with one active station ($n=1$) is 2D. That is, a single active station may have to wait as long as two times the ring latency between successive transmissions. This is because every alternate token that it receives would be unusable.

## 7 Guidelines for Setting TTRT

The FDDI standard specifies a number of rules that must be followed for setting TTRT. These rules are:

1. The token rotation time can be as long as two times the target. Thus, a synchronous station may not see the token for $2 \times T$. Therefore, *synchronous stations should request a TTRT value of one half the required* *service interval*. For example, a voice station wanting to see a token every 20 ms or less should ask for a TTRT of 10 ms.

2. *TTRT should allow at least one maximum size frame along with the synchronous time allocation, if any.* That is:

$$\begin{aligned} \text{TTRT} \geq\ & \text{Ring Latency} + \text{Token Time} \\ & + \text{Max frame time} \\ & + \text{Synchronous allocation} \end{aligned}$$

   The maximum size frame on FDDI is 4500 bytes (0.360 ms). The maximum ring latency is 1.773 ms. The token time (11 bytes including 8 bytes of preamble) is 0.00088 ms. This rule, therefore, prohibits setting the TTRT at less than 2.13 ms plus the synchronous allocation.

   Violating this rule, for example, by over allocating the synchronous bandwidth, results in unfairness and starvation.

3. *No station should request a TTRT less than $T_{min}$*, which is a station parameter. The default maximum value of T_min is 4 ms. Assuming that there is at least one station with T_min=4 ms, the TTRT on a ring should not be less than 4 ms.

4. *No station should request a TTRT more than $T_{max}$*, which is another station parameter. The default minimum value of T_max is 165 ms. Assuming that there is at least one station with T_max=165 ms, the TTRT on a ring cannot be more than this value. (In practice, many stations will use a value of $2^{22} \times 40$ ns = 167.77216 ms, which can be conveniently derived from the symbol clock using a 22-bit counter.)

In addition to these rules, the TTRT values should be chosen to allow high-performance operation of the ring. These performance considerations are now discussed.

Figure 1 shows a plot of efficiency as a function of TTRT. Three different configurations called "Typical," "Big," and "Largest" are shown.

The "Typical" configuration consists of 20 single attachment stations (**SASs**) on a 4 km fiber ring. The numbers used are based on an intuitive feeling of what a typical ring would look like and not based on any survey of actual installations. Twenty offices located on a 50 m×50 m floor would require a 2 km cable or a 4 km fiber.

The "Big" configuration consists of 100 SASs on a 200 km fiber. Putting too many stations on a single ring increases the probability of bit errors [7]. This configuration is assumed to represent a reasonably large ring with acceptable reliability.

The "Largest" configuration consists of 500 dual-attachment stations (**DASs**) a ring that is assumed to have wrapped. A DAS can have one or two media access controllers (**MACs**). In the "Largest" configuration, each DAS is assumed to have two MACs. Thus, the LAN consists of 1000 MACs in a single logical ring. This is the largest number of MACs allowed on an FDDI. Exceeding this number would require recomputation of all default parameters specified in the standard.

Figure 1 shows that for all configurations, the efficiency increases as the TTRT increases. At TTRT values close to the ring latency, the efficiency is very low, and it increases as the TTRT increases. This is one reason why the minimum allowed TTRT on FDDI T_min is 4 ms. This may lead some to the conclusion that the chosen TTRT should be chosen as large as possible. However, notice also that the gain in efficiency by increasing the TTRT (that is, the slope of the efficiency curve) decreases as the TTRT increases. The "knee" of the curve depends upon the ring configuration. For larger configurations, the knee occurs at larger TTRT values. Even for the "Largest" configuration, the knee occurs in the 6 to 10 ms range. For the "Typical" configuration, the TTRT has very little effect on efficiency as long as the TTRT is in the allowed range of 4 ms to 165 ms.

Figure 2 shows the maximum access delay as a function of the TTRT for the three configurations. In order to show the complete range of possibilities, a semi-log graph was used. The vertical scale is logarithmic while the horizontal scale is linear. The figure shows that increasing TTRT increases the maximum access delay for all three configurations. On the largest ring, using a TTRT of 165 ms would cause a maximum access delay as long as 165 seconds. This means that in a worst situation a station on such a ring may have to wait a few minutes to get a usable token. For many applications, this could be considered unacceptable, therefore, a smaller number of stations or a smaller TTRT may be preferable.

Response time will now be considered. Figure 3 shows the average response time as a function of the TTRT. The WC workload was simulated at three different load levels: 28%, 58%, and 90%. Two of the three curves are horizontal straight lines indicating that TTRT has no effect on the response times at these loads. It is only at a heavy load that the TTRT makes a difference. In fact, it is only near the usable bandwidth that TTRT has any effect on the response time. The summary of the results presented so far is that if the FDDI load is below saturation, TTRT has little effect. At saturation, a larger value of TTRT gives larger usable bandwidth, but it also results in larger access delays. Selection of TTRT requires a tradeoff between these two requirements. To allow for this tradeoff, two performance metrics are listed in Table 1 for the three configurations. A number of TTRT values in the allowed range of 4 ms to 165 ms are shown. It can be seen that a very small value such as 4 ms is undesirable since it gives poor efficiency (60%) on the "Largest" ring. A very large value such as 165 ms is also undesirable since it gives long access delays. The 8 ms value is the most desirable one since it gives 80% or more efficiency on all configurations and results in a less than 1 second maximum access delay on "Big" rings. This is, therefore, the recommended default TTRT.

Table 1: Maximum Access Delay and Efficiency as a Function of TTRT

| TTRT | Max. Access Time in Secs | | | Percent Efficiency | | |
|---|---|---|---|---|---|---|
| ms | Typical 20 SAS 4 km | Big 100 SAS 200 km | Largest 500 DAS 200 km | Typical 20 SAS 4 km | Big 100 SAS 200 km | Largest 500 DAS 200 km |
| 4 | 0.08 | 0.40 | 4.00 | 98.94 | 71.87 | 49.55 |
| 8 | 0.15 | 0.79 | 8.00 | 99.47 | 85.92 | 74.77 |
| 12 | 0.23 | 1.19 | 11.99 | 99.65 | 90.61 | 83.18 |
| 16 | 0.30 | 1.59 | 15.99 | 99.74 | 92.95 | 87.38 |
| 20 | 0.38 | 1.98 | 19.98 | 99.79 | 94.36 | 89.91 |
| 165 | 3.14 | 16.34 | 164.84 | 99.97 | 99.32 | 98.78 |

There are a few additional reasons for prefering 8 ms TTRT over a large TTRT (such as 165 ms). First, with a large TTRT a station may receive a large number of frames back-to-back. To be able to operate in such environments, adapters should be designed with large receive buffers. Although, the memory isn't considered an expensive part of computers, its cost is still a significant part of low cost components such as adapters. Also, the board space for this additional memory is significant as are the bus holding times required for such large back-to-back transfers.

Second, very large TTRT results essentially in an exhaustive service discipline, which has several known drawbacks. For example, exhaustive service is unfair. Frames coming to higher load stations have a higher chance of finding the token there in the same transmission opportunity, while frames arriving at low load stations may have to wait. Thus, the response time depends upon the load - lower for higher-load stations and higher for lower-load stations [8]. Third, the exhaustive service makes the response time of a station depend upon its location with respect to that of high load stations. The station immediately down stream from a high load station may get better service than the one immediately upstream

# 8 Effect of Extent

The total length of the fiber is called the extent of the ring. The maximum allowed extent on FDDI is 200 km. Figures 4 and 5 show the efficiency and maximum access delay as a function of the extent. A star-shaped ring with all stations at a fixed radius from the wiring closet is assumed. The total cable length, shown along the horizontal axis, is calculated as 2×Radius×Number of stations. From the figures, it can be seen that larger rings have a slightly lower efficiency and longer access delay.[4] In all cases, the performance (with TTRT=8 ms) is acceptable.

# 9 Effect of the Total Number of Stations

The total number of stations includes active as well as inactive stations. In general, increasing the number of stations increases the ring latency due to increasing fiber length and increasing the sum of station delays. Thus, the effect is similar to that of the extent. That is, a larger number of stations on one ring results in a lower efficiency and longer access delay. Another problem with a larger number of stations on a ring is the increased bit-error rate. Once again, it is preferable not to construct very large rings.

[4] The increase on access delay is not visible due to the logarithmic scale on the vertical axis.

## 10 Effect of the Number of Active Stations

As the number of active stations (or MACs) increases, the total load on the ring increases. Figures 6 and 7 show the ring performance as a function of the active number of MACs on the ring. A maximum size ring with a TTRT value of 8 ms is used. The figures show that a larger number of active MACs on a ring results in a better efficiency and a longer access delay. It is, therefore, preferable to segregate active stations on separate rings.

## 11 Effect of Frame Size

It is interesting to note that frame size does not appear in the simple models of efficiency and access delays beacuse frame size has little impact on FDDI performance. In this analysis, no "asynchronous overflow" is assumed, that is, the transmission stops instantly as the THT expires. Actually, the stations are allowed to finish the transmission of the last frame. The extra time used by a station after THT expiry is called *asynchronous overflow*. Assuming all frames are of fixed size, let $F$ denote the frame transmission time. On every transmission opportunity an active station can transmit as many as $k$ frames:

$$k = \left\lceil \frac{T - D}{F} \right\rceil$$

Here, $\lceil\rceil$ is used to denote rounding up to the next integer value. The transmission time is $kF$, which is slightly more than $T - D$. With asynchronous overflow, the modified efficiency and access delay formulae become:

$$\text{Efficiency} = \frac{nkF}{n(kF+D)+D}$$

$$\text{Access delay} = (n-1)(kF+D)+2D$$

Notice that substituting $kF = T - D$ in the above equations results in the same formulae as in Equations 1 and 2.

Figures 8 and 9 show the efficiency and access delay as function of frame size. Frame size has only a slight effect on these metrics. In practice, larger frame sizes also have the following effects:

1. The probability of error in a larger frame is larger.
2. Since the size of protocol headers and trailers is fixed, larger frames cause less protocol overhead.
3. The time to process a frame increases only slightly with the size of the frame. A larger frame size results in fewer frames and, hence, in less processing at the host.

Overall, we recommend using as large a frame size as the reliability considerations allow.

## 12 Summary

The Target Token Rotation time (TTRT) is the key network parameter that network managers can use to optimize the performance of their FDDI ring network. Other parameters that affect the performance are extent (length of cable), total number of stations, number of active stations, and frame size.

The response time is not significantly affected by the TTRT value unless the load is near saturation. Under very heavy load, response time is not a suitable metric. Instead, maximum access delay, the time between wanting to transmit and receiving a token, is more meaningful.

A larger value of TTRT improves the efficiency, but it also increases the maximum access delay. A good tradeoff is provided by setting TTRT at 8 ms. Since this value provides good performance for all ranges of configurations, it is recommended that *the default value of TTRT be set at 8 ms.*

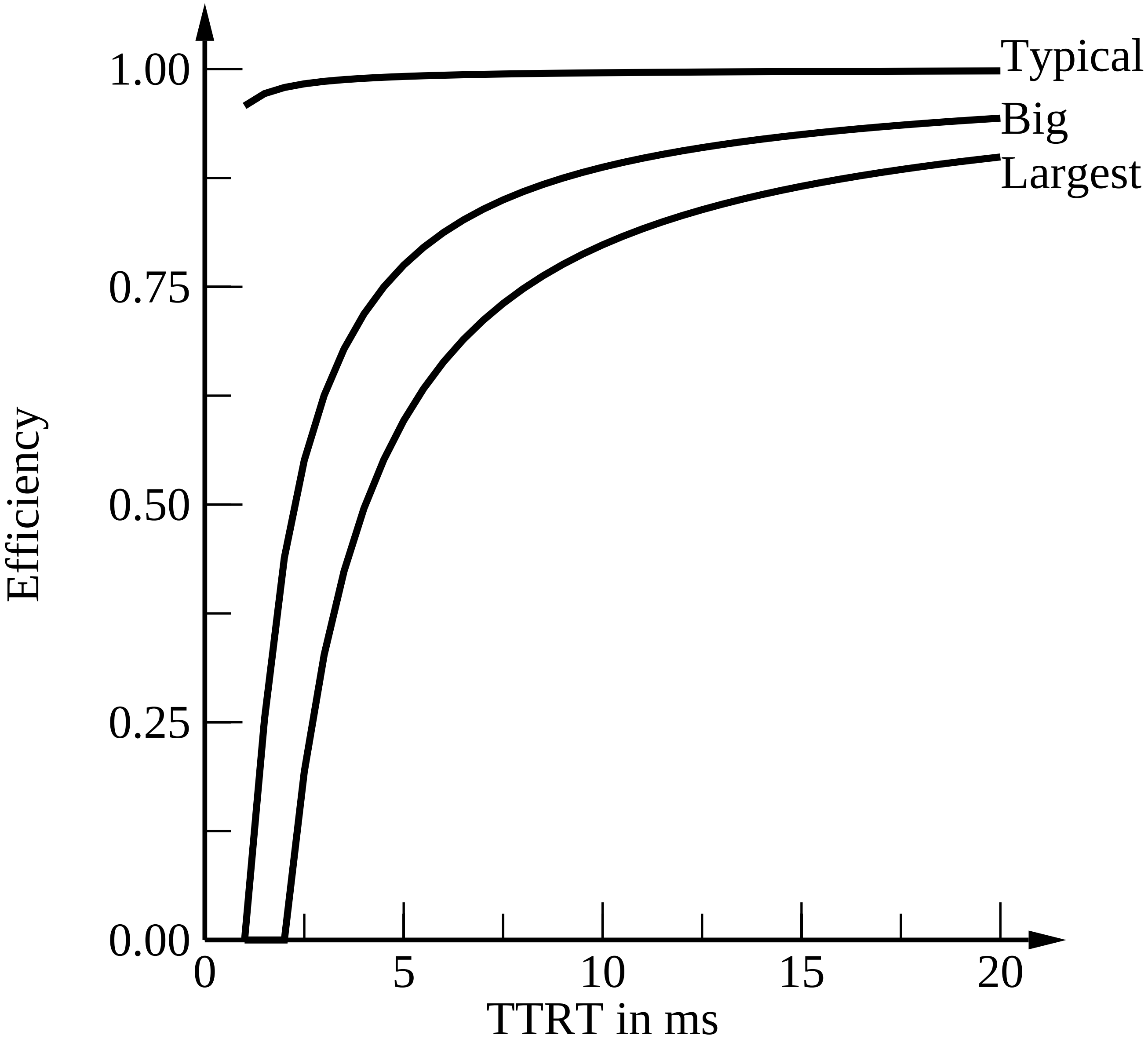

Figure 1: Efficiency as a function of the TTRT.

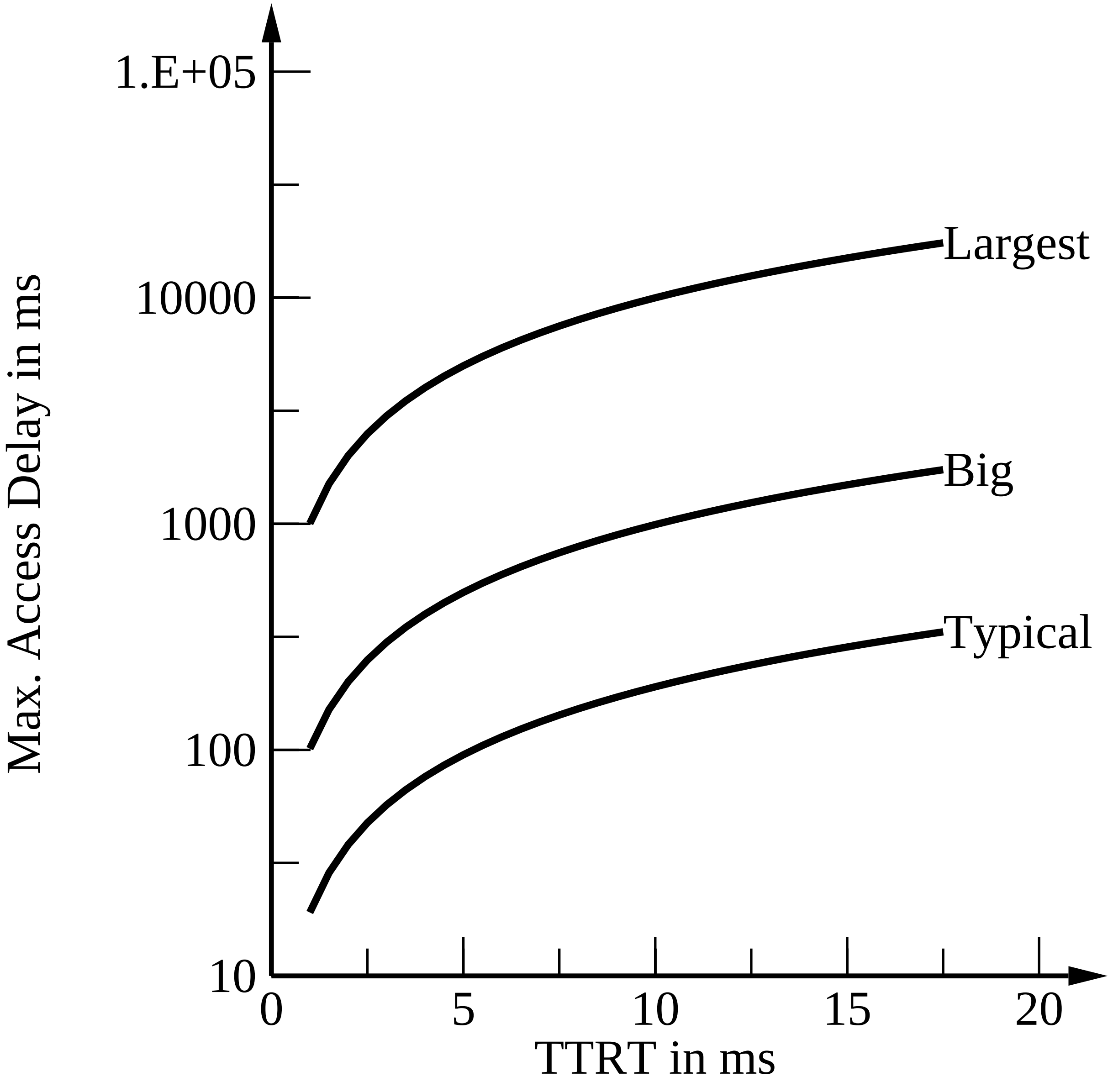

Figure 2: Access delay as a function of the TTRT.

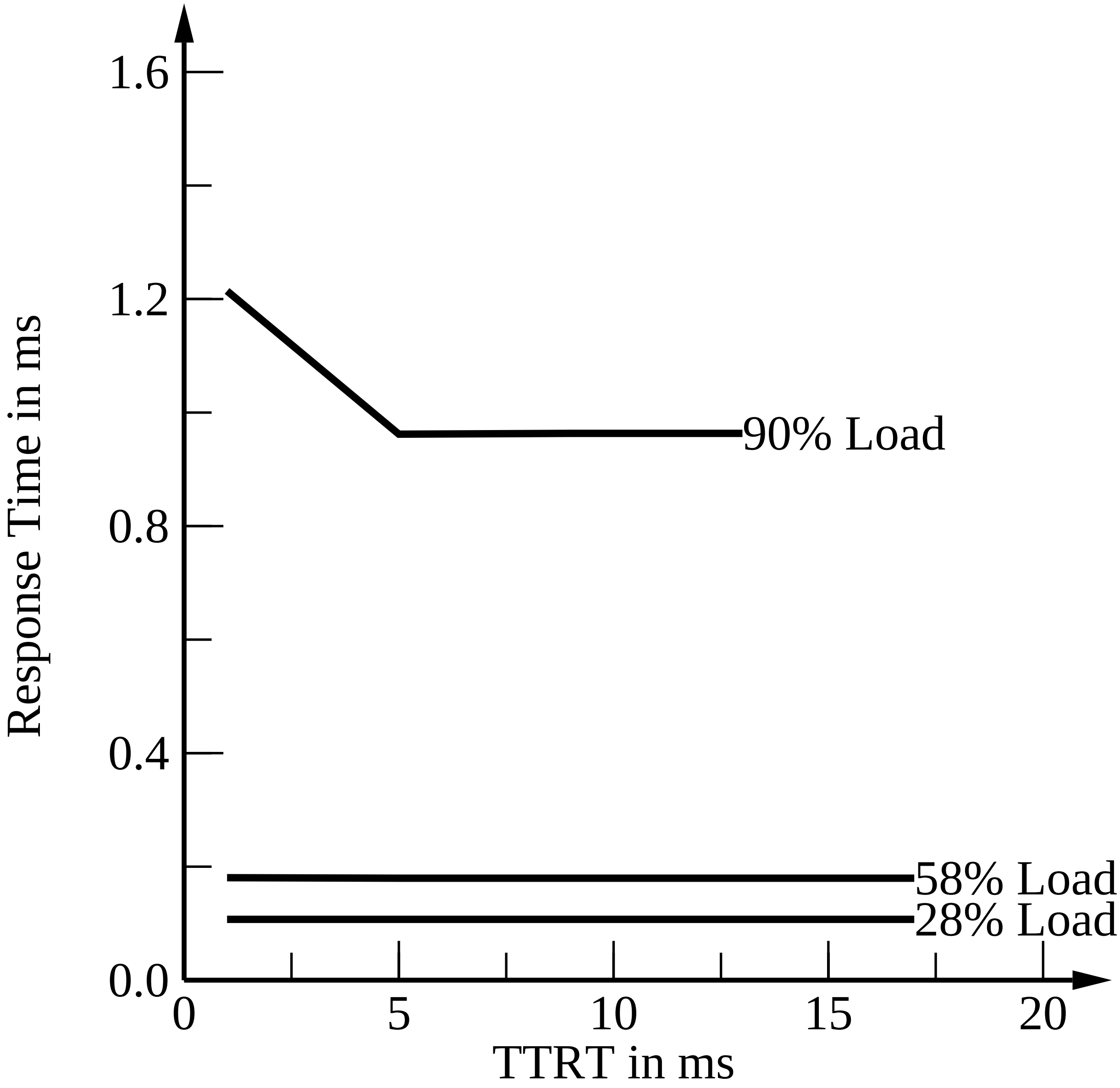

Figure 3: Response time as a function of TTRT.

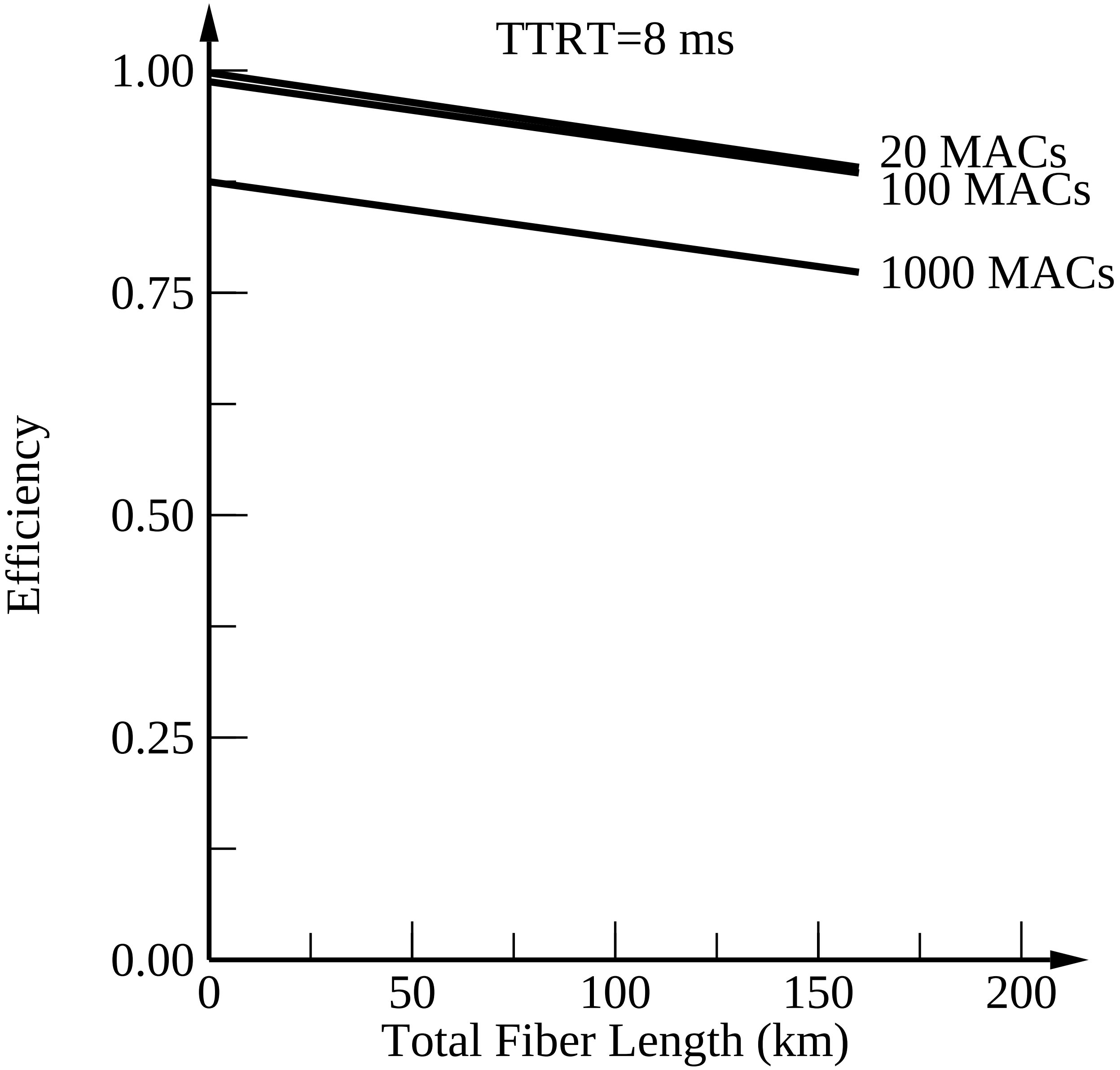

Figure 4: Efficiency as a function of the extent of the ring.

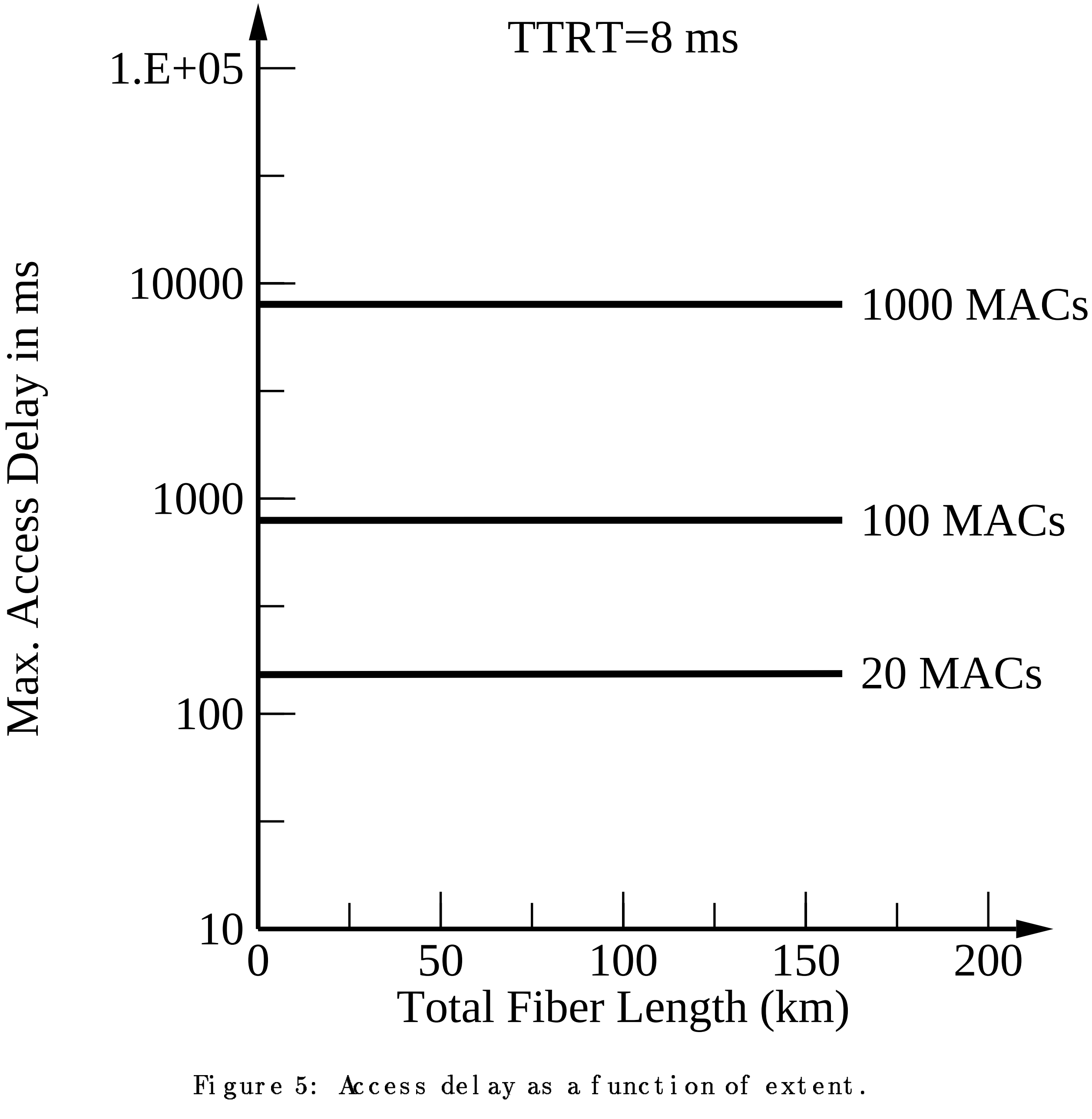

Figure 5: Access delay as a function of extent.

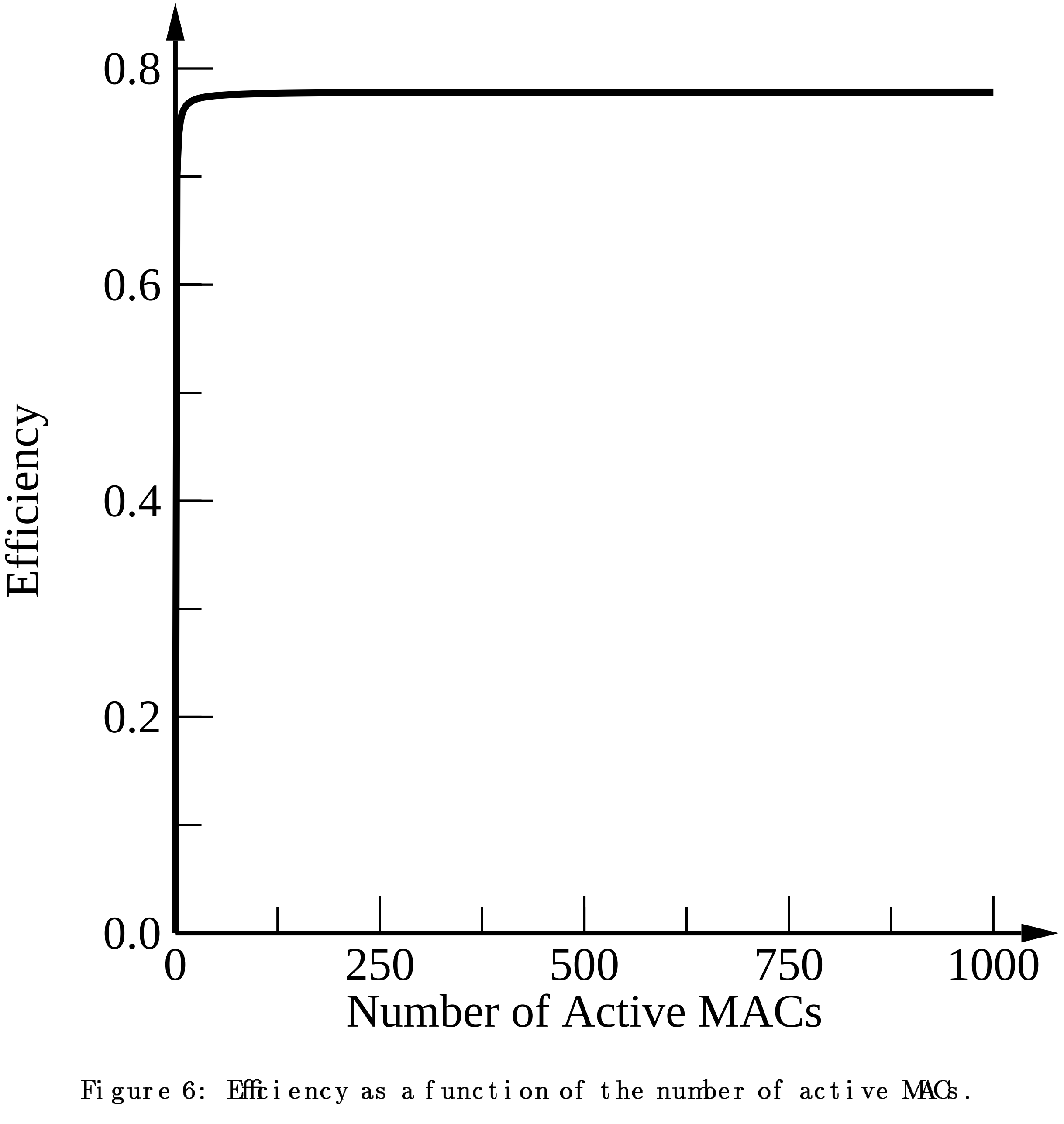

Figure 6: Efficiency as a function of the number of active MACs.

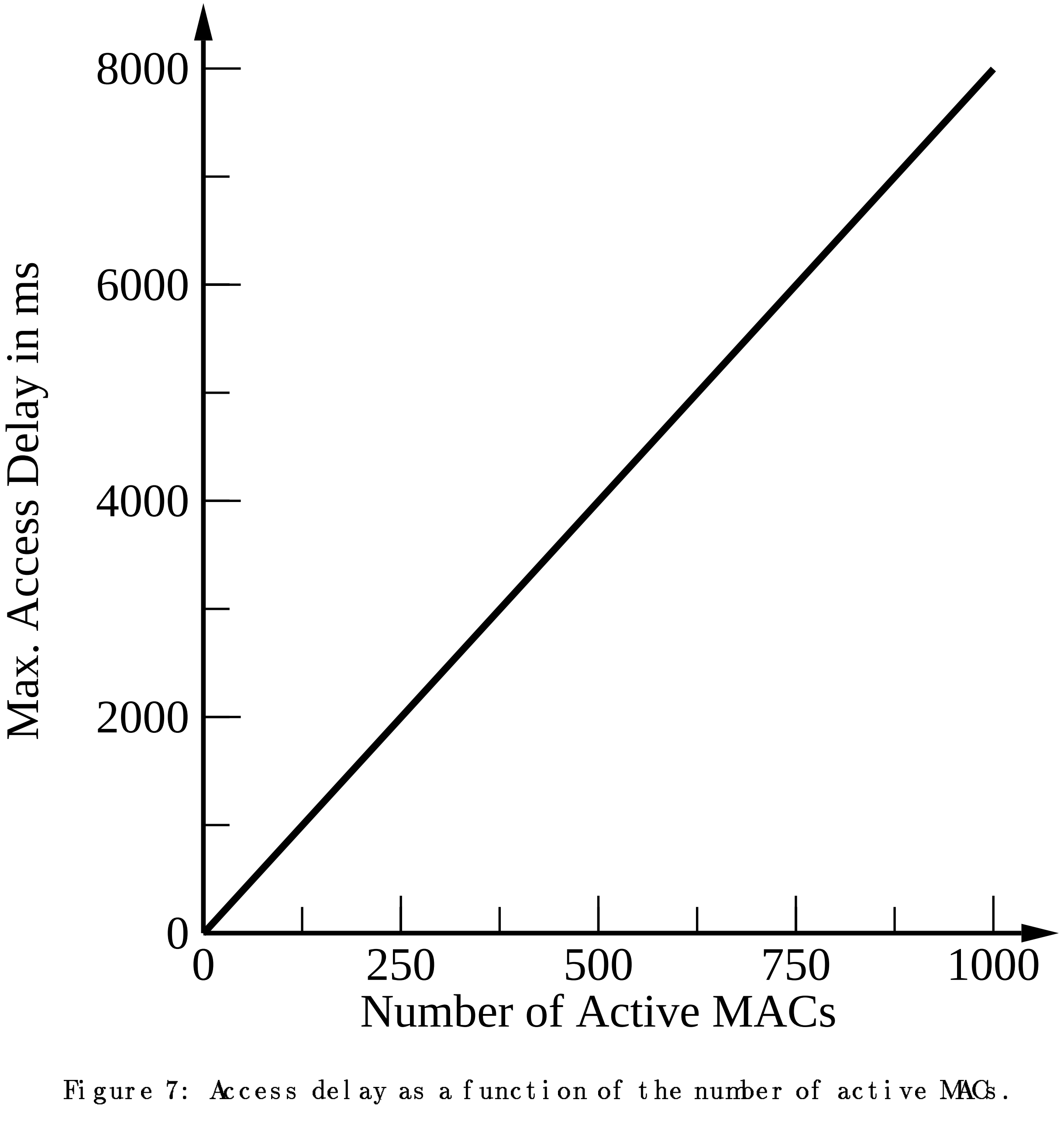

Figure 7: Access delay as a function of the number of active MACs.

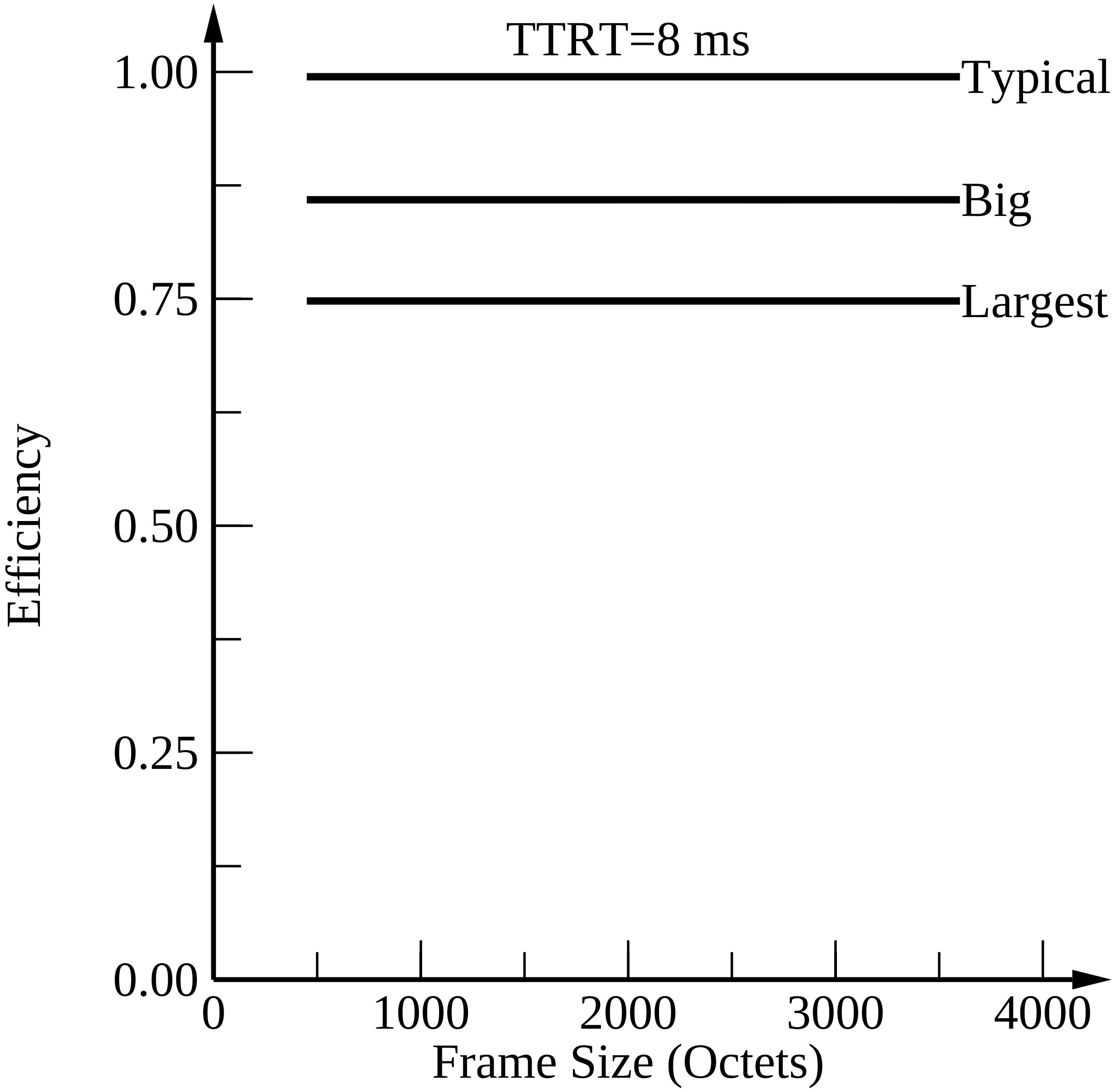

Figure 8: Efficiency as a function of the frame size.

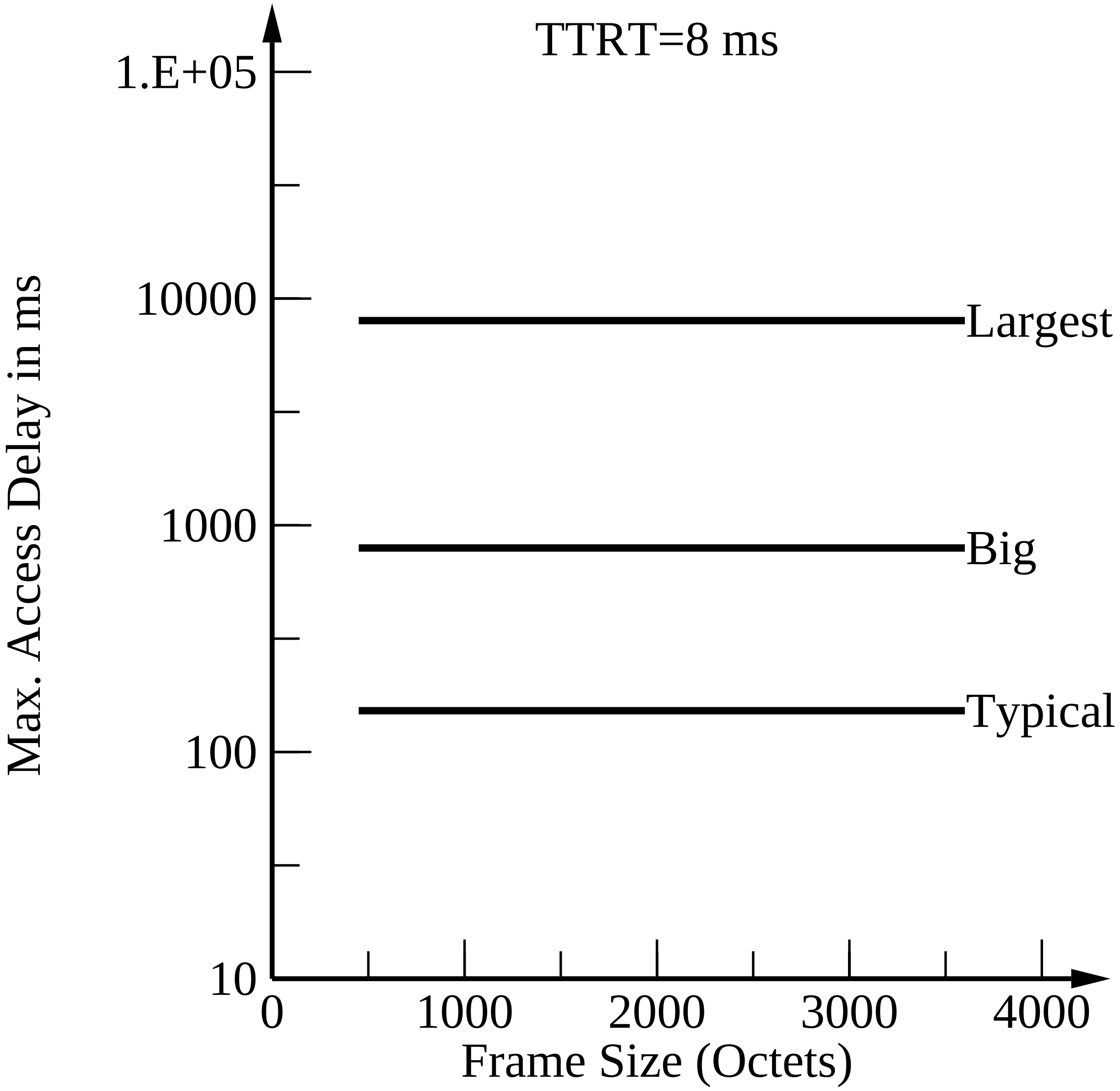

Figure 9: Access delay as a function of frame size.

## Biography

Raj Jain is a Senior Consulting Engineer at Digital Equipment Corporation where he has been involved in performance modeling and analysis of a number of computer systems and networks including VAX clusters, DECnet, Ethernet, and FDDI. He holds a Ph. D. degree from Harvard University, has sixteen years of experience in performance analysis, and has taught graduate courses in performance analysis at Massachusetts Institute of Technology. He is the author of "The Art of Computer Systems Performance Analysis" published recently by Wiley.

Raj Jain is known for introducing the packet train model of computer network traffic along with several congestion control and avoidance schemes. He has authored numerous papers on networking performance and is a popular speaker. He has actively participated as a program committee member, invited speaker, or a tutorial speaker at several conferences including SIGCOMM, SIGMETRICS, Data Communications Symposium, InterOp, ICCD, ICC, TENCON, LCN, and ITC. He was a member of a US Computer Communications delegation to China in 1987.

Raj Jain is a Senior Member of the IEEE, a member of ACM, Mathematical Association of America, Society for Computer Simulation, Operations Research Society of America, and National Writers Club. He is listed in Who's Who in the Computer Industry, 1989.